\documentclass[prb,twocolumn,preprintnumbers,amsmath,amssymb]{revtex4}

\usepackage{amsmath,latexsym,epsf,epsfig}

\newcommand{\II}{\mbox{${\mathbb I}$}}

\newcommand{\RR}{\mbox{${\mathbb R}$}}

\def\G{\mathbb G}
\def\UU{\mathbb U}
\def\S{\mathbb S}

\newcommand{\rd}{{\rm d}}
\newcommand{\diag}{{\rm diag}}

\newcommand{\U}{{\cal U}}
\newcommand{\cP}{{\cal P}}

\newcommand{\ph}{\varphi}
\newcommand{\phd}{\widetilde{\varphi}} 
\newcommand{\phs}{\varphi^{(s)}}
\newcommand{\phb}{\varphi^{(b)}}

\newcommand{\tx}{\widetilde{x}}

\def\A{\mathcal A}

\def\I{\mathcal I}
\def\der{\partial }

\def\ri{{\rm i}}
\def\xt{{\widetilde x}}
\def\tt{{\widetilde t}}
\def\prt{{\partial}}

\def\e{{\rm e}}

\begin{document}
%\leftline{Preliminary draft}
\bigskip
\bigskip

\title{Off-critical Luttinger Junctions}
\author{Brando Bellazzini$^1$, Mihail Mintchev$^2$ and Paul Sorba$^3$}
\affiliation{
${}^1$ Institute for High Energy Phenomenology
Newman Laboratory of Elementary Particle Physics,
Cornell University, Ithaca, NY 14853, USA\\
${}^2$ Istituto Nazionale di Fisica Nucleare and Dipartimento di Fisica dell'Universit\`a di Pisa,\\
Largo Pontecorvo 3, 56127 Pisa, Italy\\ 
${}^3$ Laboratoire de Physique Th\'eorique d'Annecy-le-Vieux, 
UMR5108, Universit\'e de Savoie, CNRS,\\  
9, Chemin de Bellevue, BP 110, F-74941 Annecy-le-Vieux Cedex, France}

\date{\today}

\begin{abstract} 

We investigate Luttinger junctions of quantum wires away from criticality. 
The one-body scattering matrix, corresponding to the off-critical boundary conditions at
the junction, admits in general antibound and/or bound states. Their contribution to
the theory is fixed by causality. The presence/absence of bound states determines the 
existence of two different regimes with inequivalent physical properties. 
A scattering matrix without bound states defines an   
isolated equilibrium system. Bound states instead drive the system away from equilibrium, 
giving raise to non-trivial incoming or outgoing energy flows in the junction.  
We derive in both regimes and in explicit form the electromagnetic conductance tensor, 
pointing out the different impact of bound and antibound states. 

\end{abstract}

\maketitle

\bigskip 

\section{Introduction} 

%\vskip 0.5truecm 
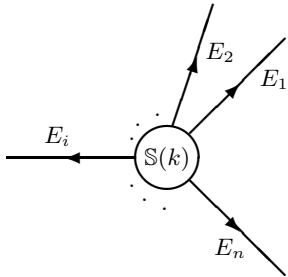
\begin{figure}[tb]
%\blankbox{.6\columnwidth}{5pc}
\setlength{\unitlength}{0.9mm}
\begin{picture}(450,20)(-15,20)
%names
\put(25.2,1.6){\makebox(20,20)[t]{$\S(k)$}}
%\put(28.5,1){\makebox(20,20)[t]{$V$}}
\put(42,11){\makebox(18,22)[t]{$E_1$}}
\put(33,17){\makebox(20,20)[t]{$E_2$}}
\put(9,3.5){\makebox(20,21)[t]{$E_i$}}
%\put(15,-1.2){\makebox(20,20)[t]{$P$}}
%\put(20,-0.8){\makebox(20,20)[t]{$x$}}
\put(34.5,-12){\makebox(20,20)[t]{$E_n$}}
%\put(33,1){\makebox(20,20)[t]{$I_1$}}
%\put(24,10){\makebox(20,20)[t]{$I_2$}}
%\put(24,-6){\makebox(20,20)[t]{$I_3$}}
%bullets
%\put(20.4,0.9){\makebox(20,20)[t]{$\bullet$}}
%\put(28.6,4.3){\makebox(20,20)[t]{$\bullet$}}
%\put(28.7,-2.7){\makebox(20,20)[t]{$\bullet$}}
%lines
\thicklines 
\put(35.2,20){\circle{10}}
\put(38.6,23.6){\line(1,1){14}}
\put(30.5,20){\line(-1,0){19}}
%\put(24.4,20){\line(-1,0){8}}
\put(38.6,16.6){\line(1,-1){14}}
\put(36,24.6){\line(1,3){6}}
%points
\put(20,3){\makebox(20,20)[t]{$.$}}
\put(20.9,5){\makebox(20,20)[t]{$.$}}
\put(23.8,6.6){\makebox(20,20)[t]{$.$}}
\put(20,-4){\makebox(20,20)[t]{$.$}}
\put(21.9,-6){\makebox(20,20)[t]{$.$}}
\put(24.8,-7.3){\makebox(20,20)[t]{$.$}}
\put(46,31){\vector(1,1){0}}
\put(46,9){\vector(1,-1){0}}
\put(39.62,35){\vector(1,3){0}}
\put(20,20){\vector(-1,0){0}}
%
%\put(15,0.7) {\makebox(20,20)[t]{$\circ$}}
\end{picture} 
\vskip 2truecm
\caption{A star graph $\Gamma$ with scattering matrix $\S(k)$ at the vertex.} 
\label{sgraph}
\end{figure}

The past two decades have shown a constantly growing 
interest\cite{kf-92,nfll-99,SS,sdm-01,mw-02,y-02,lrs-02,cte-02,ppil-03,coa-03,
dgst-03,rs-04,kd-05,klvf-05,gs-05,
emabms-05,ff-05,drs-06,Bellazzini:2006jb,Bellazzini:2006kh,Bellazzini:2008mn,
hc-08,drs-08,dr-08,hkc-08,adrs-08,Bellazzini:2008cs,dr-09,Bellazzini:2008fu,Ines,Bellazzini:2009nk,rhca-10} in 
the physics of quantum wire junctions. 
The quantum nature of the transport properties of these devises is fairly well 
described by the Tomonaga-Luttinger model\cite{h-81} on graphs 
of the type shown in FIG. \ref{sgraph}. 
The edges $\{E_i\, :\, i=1,...,n\}$ of such a {\it star} graph $\Gamma$ are 
modeling the wires, whereas the vertex represents 
the junction, which can be treated\cite{Bellazzini:2006jb} as a point-like defect (impurity) in the Luttinger liquid. 
The defect is implemented by a nontrivial one-body scattering matrix $\S(k)$, $k$ being the momentum.  
At criticality the boundary conditions at the vertex are scale invariant and the scattering matrix 
is a constant $n\times n$ matrix $\S$. Various authors\cite{coa-03,Bellazzini:2006kh,drs-06,hc-08,rhca-10} 
have investigated in this regime the physical properties of the system. In particular, they 
derived the conductance tensor $\G$, getting the following simple expression 
\begin{equation} 
\G_{ij} = \G_{\rm line} (\delta_{ij} - \S_{ij}) \, , 
\label{C1} 
\end{equation} 
the coefficient $\G_{\rm line}$ being the conductance of a single wire without junctions, which depends 
on the parameters (see eq. ({\ref{Gline}) below) of the Luttinger liquid. 

In the present paper we pursue further the study of off-critical Luttinger junctions started in 
Ref. \onlinecite{Bellazzini:2008fu}. Our main goal here is to 
extend the analysis to the case when $\S(k)$ admits 
bound states and to derive the off-critical generalization (see eq. (\ref{C2}) below) 
of the formula (\ref{C1}). This is not a purely
theoretical problem, because realistic quantum wire junctions are usually noncritical. In
order to reproduce this situation as close as possible, in what follows we will keep the
theory critical in the  {\it bulk} of $\Gamma$, allowing for breaking of scale
invariance only at the vertex.  In this regime the scattering matrix is no longer
constant, but depends on the momentum $k$.  The continuation of $\S(k)$ in the complex
$k$-plane is a meromorphic function with poles of the type 
$k=\ri \eta$ with real $\eta \not= 0$. It turns out that the behavior 
of the theory depends in a crucial way on the sign of $\eta$. 
In absence of bound states (all $\eta <0$), the model has\cite{Bellazzini:2008fu} 
a unitary time evolution respecting time-translation invariance. 
Accordingly, the energy is conserved. 
The situation radically changes when $\S(k)$ admits bound states (some $\eta >0$). 
Each of them generates in the spectrum of the theory a kind of damped harmonic oscillator. 
These oscillators lead to a breakdown of time-translation invariance. The energy of the system is no 
longer conserved, which signals a nontrivial energy flow through the boundary 
(the vertex of $\Gamma$). The essential point here is that the relative non-equilibrium 
state is fixed by the fundamental physical principle of causality (local commutativity). 
Physically, the off-critical boundary conditions with $\eta >0$ 
generate a specific boundary interaction with the environment. One possibility to implement 
such interaction in realistic quantum wire junctions might be\cite{ppil-03,coa-03,gs-05} 
an external magnetic field crossing the junction. 

The paper is organized as follows. In the next section we briefly 
review the Tomonaga-Luttinger (TL) model 
on a star graph with off-critical boundary conditions at the vertex.  
The associated scattering matrix $\S(k)$ and its analytic properties 
are described in detail. The symmetry content of the model is also 
analyzed here. In section III we examine the impact of the analytic structure of $\S(k)$
on the physical properties of the theory. The generalization of the conductance formula
(\ref{C1}) away from criticality is derived as well. The basic features of the
off-critical conductance are discussed in detail, confirming the different role played by
the bound and the antibound  states of $\S(k)$. Section IV is devoted to our conclusions.
The Appendix \ref{appA} contains some technical details concerning the quantization on 
$\Gamma$ with off-critical boundary conditions.

\section{Luttinger liquid with off-critical boundary conditions on $\Gamma$} 

\subsection{Bulk theory and boundary conditions}  

The dynamics of the Luttinger liquid in the bulk is defined by Lagrangian density 
\begin{multline}
{\cal L} = \ri \psi_1^*(\der_t - v_F\der_x)\psi_1 +  \ri \psi_2^*(\der_t + v_F\der_x)\psi_2\\ 
-g_+(\psi_1^* \psi_1+\psi_2^* \psi_2)^2 - g_-(\psi_1^* \psi_1-\psi_2^* \psi_2)^2 \, .\qquad   
\label{lagr}
\end{multline}
Here $\{\psi_\alpha (t,x,i)\,:\, \alpha =1,2\}$ are complex fields, depending on the time $t$ 
and the position $(x,i)$, where $x > 0$ is the distance from the vertex  
and $i=1,...,n$ labels the edge, as shown in FIG. \ref{sgraph}. Finally, $v_F$ is the Fermi velocity and 
$g_\pm \in \RR$ are the coupling constants\cite{f1}. The equations of motion following from 
(\ref{lagr}) imply the conservation law  
\begin{equation}
\der_t \rho (t,x,i) -v_F\der_x j (t,x,i)= 0\, , 
\label{conservation}
\end{equation}
where   
\begin{eqnarray}
\rho (t,x,i) &=& \left (\psi_1^*\psi_1 + \psi_2^*\psi_2 \right )(t,x,i)\, , \\
\label{density}
j (t,x,i) &=& \left (\psi_1^*\psi_1 - \psi_2^*\psi_2 \right )(t,x,i)\, , 
\label{current}
\end{eqnarray} 
are the charge density and electric current respectively. Our main task below  
is to derive the relative conductance. For this purpose we have to 
complete first the description of the dynamics, specifying the interaction 
at the vertex of $\Gamma$. In other words, we must fix the boundary conditions 
on $\psi_\alpha$ at $x=0$. This is a very delicate point because it 
strongly interferes with the solution of the model. It is useful to recall in this respect 
that the model (\ref{lagr}) is exactly solvable on the line\cite{h-81} via 
bosonization. In order to preserve this nice feature on 
$\Gamma$, it is more convenient to formulate the boundary conditions 
directly in bosonic terms. We will show now that this strategy, which 
works\cite{Bellazzini:2006kh,Bellazzini:2008mn,Bellazzini:2008fu} nicely 
at criticality, can be extended to off-critical boundary conditions as well. 

The basic ingredient for solving the model (\ref{lagr}) via bosonization is the 
scalar field $\ph$ satisfying 
\begin{equation}
\left (\prt_t^2 - \prt_x^2 \right )\ph (t,x,i) = 0\, , \qquad x>0 
\label{eqm1}
\end{equation} 
and some boundary conditions for $x=0$. Following a standard QFT procedure, 
the latter are fixed by requiring that the operator $K\equiv -\prt_x^2$ on $\Gamma$ is 
{\it self-adjoint}. This problem has been 
intensively investigated in the recent mathematical literature\cite{ks-00,qg,H1}, 
where the subject goes under the name of ``quantum graphs". From 
these studies one infers that $K$ is self-adjoint on $\Gamma$ 
if and only if the field $\ph$ satisfies the 
boundary condition\cite{ks-00,H1} 
\begin{equation} 
\sum_{j=1}^n \left [\lambda (\II-\UU)_{ij}\, \ph (t,0,j) -\ri (\II+\UU)_{ij}
(\prt_x\ph ) (t,0,j)\right ] = 0\, , 
\label{bc} 
\end{equation} 
where $U$ is any unitary matrix and $\lambda > 0$ is a 
parameter with dimension of mass, characterizing the breaking of scale invariance. 

Eq. (\ref{bc}) generalizes to the graph $\Gamma$ the familiar 
mixed (Robin) boundary condition on the half-line $\RR_+$. 
The matrices $\UU=\II$ and $\UU=-\II$ define the Neumann and Dirichlet 
boundary conditions respectively. The physical interpretation of 
(\ref{bc}) in the context of bosonization was discussed in Ref. \onlinecite{Bellazzini:2008mn}. 
At criticality (\ref{bc}) describes the splitting of the electric current (\ref{current}) 
at the junction.\cite{f2}

\subsection{Scattering matrix and solution of the model}

As already mentioned in the introduction, for the explicit construction 
of $\ph$ it is convenient to interpret \cite{Bellazzini:2006jb} the vertex of $\Gamma$ as a
point-like impurity (defect) \cite{Liguori:1996xr,Mintchev:2002zd,Mintchev:2003ue,Mintchev:2004jy}, 
characterized by a non-trivial scattering matrix $\S(k)$. This matrix 
is associated \cite{ks-00,qg} to the operator $K$ on $\Gamma$
and is fully determined by the boundary conditions (\ref{bc}). The 
explicit form is \cite{ks-00}
\begin{equation} 
\S (k) = -\frac{[\lambda (\II - \UU) - k(\II+\UU )]}{[\lambda (\II - \UU) + k(\II+\UU )]}  
\label{S1}
\end{equation} 
and has a simple physical interpretation: the diagonal element $\S_{ii}(k)$ 
represents the reflection amplitude from the defect on the edge $E_i$, whereas  
$\S_{ij}(k)$ with $i\not=j$ equals the transmission amplitude from $E_i$ to $E_j$. 

The matrix (\ref{S1}) has a number of remarkable features. It is 
unitary $\S(k)^*=\S(k)^{-1}$ by construction and satisfies 
$\S(k)^*=\S(-k)$, known as Hermitian analyticity. Notice also that 
$\S(\lambda ) = \UU$, showing that the boundary condition (\ref{bc}) is fixed actually 
by the value of scattering matrix at the scale $\lambda$. 

Let us summarize now the analytic properties of $\S(k)$ needed in what follows. 
We denote by $\U$ the unitary matrix diagonalizing $\UU$ and parametrize 
\begin{equation} 
\UU_d=\U^{-1}\, \UU\, \U 
\label{d1}
\end{equation}  
as follows 
\begin{equation} 
\UU_d = \diag \left (\e^{2\ri \alpha_1}, \e^{2\ri \alpha_2}, ... , \e^{2\ri
\alpha_n}\right )\, , \qquad \alpha_i \in \RR\, . 
\label{d2}
\end{equation} 
Using (\ref{S1}), one easily verifies that $\U$ 
diagonalizes also $\S(k)$ {\it for any} $k$ and that 
\begin{multline} 
\S_d(k) = \U^{-1} \S(k) \U = \\
\diag \left (\frac{k+\ri \eta_1}{k-\ri \eta_1}, \frac{k+\ri \eta_2}{k-\ri \eta_2}, ... , \frac{k+\ri \eta_n}{k-\ri \eta_n} \right ) \, , 
\label{d3}
\end{multline} 
where 
\begin{equation} 
\eta_i = \lambda \tan (\alpha_i)\, , 
\qquad -\frac{\pi}{2} \leq \alpha_i \leq \frac{\pi}{2}\, .  
\label{p1}
\end{equation} 
We conclude therefore that $\S(k)$ is a meromorphic function with {\it simple} poles 
located on the imaginary axis and different from 0. In what follows we denote by 
$\cP = \{\ri \eta \, :\, \eta\not= 0\}$ the set of {\it distinct} poles of $\S(k)$: 
the subset $\cP_+ = \{\ri \eta \, :\, \eta>0\}$ in the upper half-plane 
corresponds to {\it bound} states whereas $\cP_- = \{\ri \eta \, :\, \eta<0\}$ 
in the lower half-plane gives raise to {\it antibound} states.\cite{f3} 
We will make use below also of the {\it residue} matrix defined by 
\begin{equation} 
R^{(\eta )}_{ij} = 
\frac{1}{\ri \eta }\, \lim_{k\to \ri \eta } (k-\ri \eta )\S_{ij}(k)\, , \qquad
\ri\eta \in \cP  \, . 
\label{residuem}
\end{equation} 

We recall that in the above parametrization the angular variables $\alpha_i$ 
characterize the departure from criticality, the critical points corresponding\cite{Bellazzini:2008fu} 
to the values $\alpha_i = 0, \pm \pi/2$. The classification and the stability 
properties of these critical points have been extensively studied in the 
literature\cite{nfll-99,lrs-02,coa-03,drs-06,Bellazzini:2006kh,Bellazzini:2008mn,Bellazzini:2008fu,hc-08,Bellazzini:2009nk}. 
As already stated in the introduction, our goal here is to go beyond and explore the theory away from criticality.  

The basic steps in the construction of the field $\ph$, which satisfies 
the wave equation (\ref{eqm1}) and the off-critical boundary condition (\ref{bc}), 
are given in Appendix \ref{appA}. The subtle point 
is the contribution of the bound states of $\S(k)$, which need a 
separate treatment where local commutativity turns out to be essential. 
The dual field $\phd$ is defined by 
\begin{align}
\partial_{t}\phd(t,x,i)=&-\partial_{x}\ph(t,x,i)\, ,\\ 
\partial_{x}\phd(t,x,i)=&-\partial_{t}\ph(t,x,i)\, ,
\label{D1}
\end{align} 
and the solution of the Luttinger model 
on $\Gamma$ can be expressed in terms of the pair $\{\ph,\, \phd\}$. For the details we 
refer to Refs. \onlinecite{Bellazzini:2006kh,Bellazzini:2008mn,Bellazzini:2008fu}, 
recalling here only the solution 
\begin{eqnarray}  
\psi_1(t,x,i) &\sim&
:\e^{\ri \sqrt {\pi} \left [\zeta_+ \ph (vt, x,i) - \zeta_- \phd (vt, x,i)\right ]}:\,, 
\label{psi1}\\
\psi_2(t,x,i) &\sim&
:\e^{\ri \sqrt {\pi} \left [\zeta_+\ph (vt, x, i) + \zeta_- \phd (vt+x)\right ]}:\,, 
\label{psi2}
\end{eqnarray} 
and the expression of the electric current (\ref{current}) 
\begin{equation} 
j (t,x,i) = - \frac{v}{v_F \zeta_+ \sqrt {\pi }}\, \der_x \ph (vt, x, i) \, ,  
\label{bcurrent}
\end{equation} 
needed in the derivation of the conductance. 
In (\ref{psi1}, \ref{psi2}) $: \cdots :$ denotes the normal product, whereas  
\begin{eqnarray}
\zeta_\pm &=& \sqrt{|\kappa|} \left (
\frac{\pi \kappa v_F+2g_+}{\pi \kappa v_F+2g_-}\right )^{\pm \frac{1}{4}}\, , 
\label{z}\\ 
v&=&\frac{\sqrt{(\pi \kappa v_F+2g_-)(\pi \kappa v_F+2g_+)}}{\pi|\kappa|}\, , 
\label{v}
\end{eqnarray} 
$\kappa$ being the {\it statistical parameter}. 
The conventional fermionic Luttinger liquid 
is obtained for $\kappa =1$. For $\kappa \not=1$ one has {\it anyonic} 
Luttinger liquids. Notice also that the above solution of the Luttinger 
model is meaningful for coupling constants satisfying $2g_\pm >-\pi \kappa v_F$. 
\bigskip 

\subsection{Symmetry content} 

The bulk theory, defined by the Lagrangian (\ref{lagr}), is invariant under 
time reversal and global $U(1)$ gauge transformations. Attempting to lift 
these symmetries to theory on whole star graph $\Gamma$, 
one gets some restrictions on the boundary conditions (\ref{bc}). Time reversal 
symmetry implies\cite{Bellazzini:2009nk} that $\UU$ must be symmetric, 
\begin{equation}
\UU^t = \UU \, .
\label{T1}
\end{equation} 
Concerning the conservation of the $U(1)$ (electric) charge 
\begin{equation} 
Q= \sum_{i=1}^n \int_0^\infty \rd x\, j(t,x,i) 
\label{echarge}
\end{equation} 
on $\Gamma$, one knows already from classical electrodynamics that the Kirchhoff's rule 
\begin{equation} 
\sum_{i=1}^n j(t,0,i)= 0 
\label{K1}
\end{equation} 
must be satisfied at the vertex of $\Gamma$. Using (\ref{bcurrent}),  one can 
verify\cite{Bellazzini:2006kh,Bellazzini:2008mn,Bellazzini:2008fu} that (\ref{K1}) 
holds if and only if 
\begin{equation} 
\sum_{i=1}^{n}\S_{ij}(k)=1\, , \quad  \forall j =1,...,n\, ,  
\label{K2} 
\end{equation} 
or, equivalently 
\begin{equation} 
\sum_{i=1}^{n}\UU_{ij} = 1\, , \quad  \forall j =1,...,n\, .  
\label{K3} 
\end{equation} 
Combining (\ref{residuem}) and (\ref{K2}), one gets 
\begin{equation} 
\sum_{i=1}^n R^{(\eta )}_{ij} = 0\, , \qquad
\forall \; \ri\eta \in \cP  \, , 
\label{K4}
\end{equation} 
which will be essential below for checking the Kirchhoff's rule for the conductance away from 
criticality. 

In what follows we assume that both (\ref{T1}) and (\ref{K3}) hold. The case with broken 
time reversal has been analyzed recently in Ref. \onlinecite{Bellazzini:2009nk}. 

\section{Basic features of the model away from criticality} 

\subsection{Impact of the analytic structure of $\S(k)$} 

The analytic properties of the scattering matrix $\S(k)$ deeply influence the 
physics of the Luttinger liquid on $\Gamma$. 
The simplest observable, one can investigate in order to illustrate this fact, is the 
electric current $j$. More precisely, it is enough to study the relative two-point function, 
which is fixed\cite{Bellazzini:2008cs} up to a real parameter $\tau$ by causality (local commutativity). 
Postponing the discussion of the physical meaning of $\tau$ to 
the end of this subsection, we consider first the current-current correlator following 
from the definition (\ref{bcurrent}). In the Fock representation of the field $\ph$, defined 
by equations (\ref{dec1}), (\ref{sol1}) and (\ref{bfield}) in the Appendix \ref{appA}, one finds 
\begin{widetext} 
\begin{equation} 
\langle j(t_1,x_1,i_1)  j(t_2,x_2,i_2) \rangle = 
\frac{v^2}{\left (2\pi \zeta_+v_f \right )^2}\left [ D_{i_1i_2}(vt_{12},x_1,x_2) + 
A_{i_1i_2}(vt_{12},x_1,x_2) + B_{i_1i_2}(vt_1,vt_2,x_1,x_2;\tau )\right ] \, , 
\label{jj}
\end{equation}
where 
\begin{equation}
D_{i_1i_2}(t,x_1,x_2) = -\delta_{i_1i_2} \left [d^2(t-x_{12}) + d^2(t-x_{12}) +
d^2(t+\xt_{12}) +d^2(t-\xt_{12})\right ]\, , 
\label{fd1}
\end{equation} 
\begin{multline}
A_{i_1i_2}(t,x_1,x_2) = \sum_{\ri \eta \in \cP_-} R^{(\eta)}_{i_1i_2}\bigl \{d^2(t+\xt_{12}) + d^2(t-\xt_{12}) 
+\eta \left [d(t+\xt_{12}) -d(t-\xt_{12})\right ] \\ - 
\eta^2\left [w_-(-\eta(t-\xt_{12}) +w_+\left (\eta(t+\xt_{12})\right ) \right ]\bigr \}\, , 
\label{fa1}
\end{multline}
\begin{multline}
B_{i_1i_2}(t_1,t_2,x_1,x_2;\tau) = \sum_{\ri \eta \in \cP_+} R^{(\eta)}_{i_1i_2}
\bigl \{d^2(t_{12}+\xt_{12}) + d^2(t-\xt_{12}) 
+\eta \left [d(t_{12}+\xt_{12}) -d(t_{12}-\xt_{12})\right ] \\ - 
\eta^2\left [w_+(-\eta(t_{12}-\xt_{12}) +w_-\left (\eta(t_{12}+\xt_{12})\right ) \right ]
+2\eta^2\e^{-\eta \xt_{12}}\left [\cosh(\eta (\tt_{12}-2\tau)) -\ri \sinh (\eta t_{12})\right ]  \bigr \}\, . 
\label{fb1}
\end{multline}
\end{widetext} 
Here and in what follows $t_{12}= t_1-t_2$, $x_{12}= x_1-x_2$, 
$\tt_{12}= t_1+t_2$, $\xt_{12}= x_1+x_2$ and 
\begin{equation} 
d (\xi ) = \frac{1}{\xi - i\epsilon } \, , \qquad 
w_{\pm}(\xi)=\e^{-\xi}\, {\rm Ei} (\xi \pm i\epsilon ) \, , 
\label{dw}
\end{equation} 
with $\epsilon > 0$ and $\rm Ei$ the exponential integral function. 
By construction the functions $A_{i_1i_2}$ and $B_{i_1i_2}$ collect 
the contributions of the antibound and the bound states respectively. 
The main feature distinguishing these two functions is their time dependence. 
Notice indeed that $A_{i_1i_2}$ depends exclusively on $t_{12}$, whereas $B_{i_1i_2}$ 
depends in addition on $\tt_{12}$ and on the parameter $\tau$. 
This fact suggest the consideration of two separate cases: 

(i) When bound states are absent ($\cP_+ = \emptyset $), the behavior of the system 
is quite orthodox; the function $B_{i_1i_2}$ vanishes identically 
and time translations invariance is preserved. Accordingly, the energy 
is conserved. Because of (\ref{T1}), the theory is also invariant under time reversal 
\begin{equation}
t \rightarrow -t \, . 
\label{str}
\end{equation} 

(ii) When bound states are present ($\cP_+ \not= \emptyset $) the situation 
changes drastically. Besides on $t_{12}$, the dynamics depends also on 
$\tt_{12}$, implying that the theory is no longer invariant under time translations. 
Therefore, the energy is not conserved. The theory depends on the parameter $\tau$ 
as well and is invariant under the time reversal transformation 
\begin{equation}
t \rightarrow -t +2\tau \, , 
\label{mtr}
\end{equation} 
which is actually a reflection with respect to $\tau$. 

The properties collected in point (ii) deserve a more detailed discussion. 
They imply that for boundary conditions admitting bound states 
(i.e. some $\alpha_i \in (0\, ,\pi/2)$ - see eqs.(\ref{bc}, \ref{d1}-\ref{p1})), 
the system is not isolated. In fact, there exists a nontrivial energy flow 
crossing the boundary\cite{f4} at $x=0$, the direction 
(outgoing or incoming) being controlled by the parameter $\tau$. 
In order to illustrate this statement, let us consider the energy density $\theta $ of the TL 
model on $\Gamma$. In bosonic coordinates $\theta $ takes the form\cite{DellAntonio:1971zt} 
\begin{equation} 
\theta (t,x,i) = \frac{g_-}{\pi \zeta_-^2} (\partial_x \ph)^2(t,x,i) + 
\frac{g_+}{\pi \zeta_+^2} (\partial_x \phd\, )^2(t,x,i) \, . 
\label{ed}
\end{equation} 
As well known, the field products at coinciding points in (\ref{ed}) contain divergences 
which must be subtracted. A natural way to fix the subtraction is to take as a reference 
point the vacuum energy $\langle \theta (t,x)\rangle_{\rm line}$ on the line.\cite{f5} Using  
conventional point splitting regularization, one gets in this way\cite{Bellazzini:2008cs} 
\begin{multline} 
{\cal E} (t,x,i) = \langle \theta (t,x,i)\rangle - \langle \theta (t,x)\rangle_{\rm line} \sim \\ 
{\cal E}(x,i)+ \sum_{\ri \eta \in \cP_+} \eta^2 R^{(\eta)}_{ii} \e^{-2\eta x} \cosh [2\eta (t-\tau)] \, , 
\label{venergy}
\end{multline}
where ${\cal E}(x,i)$ is a time independent contribution,  
derived in Ref. \onlinecite{Bellazzini:2006jb}} but 
irrelevant for what follows. From (\ref{venergy}) we deduce 
that $\tau$ is the instance in which the vacuum energy flow inverts its direction, 
being outgoing for $t<\tau$ and incoming for $t>\tau$. This feature is related to the fact 
that the theory is invariant under the operation (\ref{mtr}), which obviously 
exchanges the two time intervals $(-\infty\, ,\, \tau)$ and $(\tau\, ,\, \infty )$.  
These considerations fix the physical meaning of the parameter $\tau$, 
which appears in the theory when $\cP_+ \not= \emptyset $. 

Summarizing, the behavior of off-critical Luttinger junctions is characterized by 
the two different regimes described in points (i) and (ii) above. In the case 
(i) one deals with an isolated system in equilibrium. Both energy and electric 
charge are conserved. The presence of bound states in the regime (ii) 
significantly modifies this behavior. The energy is no longer conserved 
because the relative boundary conditions determine a specific boundary 
interaction with the environment, which drives the system out of equilibrium. 
A remarkable feature is that the corresponding non-equilibrium state is 
fixed, up to the value of the parameter $\tau$, by the basic physical 
requirement of causality. 

We stress in conclusion that the above results are obtained in an abstract 
setup, where we attempt to describe the physics of a quantum wire junction 
by a Luttinger liquid with specific off-critical boundary conditions at the vertex 
of the star graph approximating the junction. Further investigations are 
needed for clarifying the plausibility of these assumptions and the applicability 
of the results to real-life quantum wire junctions. The case (i) looks 
more physical since the energy is conserved and bounded from below. 
In our opinion however, also the regime (ii) can not be excluded a priori, if one 
considers the possible interactions of the junction with the environment. In order to clarify 
the situation, one should analyze in the above framework some physical 
observables, which in principle can be checked experimentally. 
As an example, we derive in the next subsection the electromagnetic 
conductance in explicit form. Another attractive possibility to apply the 
above results is the development of an effective description of the Luttinger 
liquid on complex quantum wire networks with several junctions and loops, 
crossed possibly by magnetic fluxes implementing the interaction with the environment. 
The complete field theory analysis of such networks 
is usually a complicated problem. One approximate way to face the  
problem could be to use the star product approach \cite{KS,Schrader:2009su}, 
the ``gluing" technique \cite{Mintchev:2007qt, Ragoucy:2009hf,Caudrelier:2009ay} 
or transfer matrix formalism\cite{Khachatryan:2009xg} for deriving the {\it effective}  
scattering matrix $\S_{\rm eff}(k)$ relative to the {\it external} edges of the network. 
$\S_{\rm eff}(k)$ admits in general both bound and antibound states 
and can be used\cite{Ragoucy:2009hf} for constructing a simplified model with one effective 
off-critical junction.  

\bigskip

\subsection{Off-critical conductance} 

In order to compute the conductance, we couple 
the Luttinger liquid on $\Gamma$ to a uniform {\it classical} electric field\cite{f6} 
$E(t,i)=\partial_t A_x(t,i)$, performing in (\ref{lagr}) the substitution 
\begin{equation} 
\prt_x \longmapsto \prt_x + \ri A_x (t,i) \, .  
\label{covder}
\end{equation} 
This external field coupling deforms the bosonized version of the current according to\cite{Bellazzini:2006kh}  
\begin{multline} 
j(t,x,i) \longmapsto J(t,x,i) =  \\
- \frac{v}{v_F \zeta_+ \sqrt {\pi }}\left [ \der_x \ph (vt, x, i) +  \frac{1}{\zeta_+ \sqrt {\pi }} A_x(t,i)\right ] 
\label{Jcurrent}
\end{multline} 
The Hamiltonian encoding the interaction of $\ph$ and $A_x$ is time dependent and reads\cite{Bellazzini:2006kh} 
\begin{equation} 
H_{\rm int}(t) =  
\frac{1}{\zeta_+ \sqrt{\pi }}\sum_{i=1}^n \int_0^\infty \rd x (\prt_x \ph)(vt,x,i) A_x(t,i) \, . 
\label{hint}
\end{equation} 
We stress that now the system can exchange energy with the environment in 
two different ways. The first one corresponds to the external force 
produced by the coupling with the time dependent electric field $E(t,i)$. 
The second one represents an intrinsic property in the regime (ii), being related to the presence 
of bound states of $\S(k)$. The conductance, derived below, keeps track of both of them. 

In order to derive the conductance, one should compute the expectation value of the current (\ref{Jcurrent}) 
in the external field $A_x$. This expectation value is given by the following 
well known\cite{FW} series expansion 
\begin{widetext} 
\begin{multline}
\langle J(t,x,i)\rangle_{A_x} =  
\langle J(t,x,i) \rangle -\ri\int_{-\infty}^{t} \rd\tau_1
\langle [J(t,x,i)\, ,\, H_{\rm int}(\tau_1)]\rangle + \cdots  \\
+(-\ri)^n \int_{-\infty}^{t} \rd\tau_1 \int_{-\infty}^{\tau_1} \rd\tau_2 \cdots \int_{-\infty}^{\tau_{(n-1)}} \rd\tau_n 
[\cdots[[J(t,x,i)\, ,\, H_{\rm int}(\tau_1)]\, ,\, H_{\rm int}(\tau_2)]\, ,\cdots , H_{\rm int}(\tau_n)] + \cdots  
\label{lrt1}
\end{multline} 
in powers of $H_{\rm int}$. This expansion is particularly simple in our case. 
In fact, from (\ref{Jcurrent},\ref{hint}) one obtains that  
\begin{equation} 
[[J(t,x,i)\, ,\, H_{\rm int}(\tau_1)]\, ,\, H_{\rm int}(\tau_2)]=0\, , 
\label{dcomm}
\end{equation} 
cutting the infinite series (\ref{lrt1}) to the sum of the first two terms, namely 
\begin{equation}
\langle J(t,x,i)\rangle_{A_x} = 
\frac{v}{v_F \zeta_+^2 \pi }\left [A_x(t,i)+
\ri \sum_{j=1}^{n}\int_{-\infty}^{t} \rd\tau 
\int_0^\infty \rd y A_y(\tau,j)
\langle [\partial_y\ph(v\tau,y,j)\, ,\, \partial_x\ph(vt,x,i)]\rangle \right ]\, . 
\label{lrt2}
\end{equation} 
The expectation value $\langle J(t,x,i)\rangle_{A_x}$ is therefore linear in $A_x$, implying that the 
linear response approximation\cite{FW} to $\langle J(t,x,i)\rangle_{A_x}$ is actually exact in this case. 

Let us suppose now that the external field is switched on at 
$t=t_0$, i.e. $A_x(t,i) = 0$ for $t<t_0$. 
Using the representation (\ref{comm}) for the commutator in the left hand side of (\ref{lrt2}),  
one finds for $t >t_0$ 
\begin{equation}
\langle J(t,0,i)\rangle_{A_x} = 
G_{\rm line}\sum_{j=1}^n\int_{-\infty}^\infty \frac{\rd \omega}{2\pi} 
{\hat A}_x(\omega, j)\e^{-\ri \omega t} 
\left [\delta_{ji} - S_{ji}\left (\frac{\omega}{v} \right ) - 
\sum_{\ri \eta \in \cP} R_{ji}^{(\eta)} \frac{v \eta}{v \eta + \ri \omega} 
\e^{(t-t_0)(v \eta+\ri \omega)} \right ]\, , 
\label{lrt3}
\end{equation}
where 
\begin{equation} 
G_{\rm line} = \frac{v}{v_F \zeta_+^2 \pi }\, , \qquad 
{\hat A}_x(\omega, i) = \int_{-\infty}^\infty \rd \tau\, \e^{\ri \omega \tau} A_x(\tau,i)\, .  
\label{Gline}
\end{equation} 
The conductance can be extracted directly from (\ref{lrt3}). The result is   
\begin{equation} 
G_{ij}(\omega, t-t_0) =  
G_{\rm line} \left [\delta_{ij} - S_{ij}\left (\frac{\omega}{v} \right ) - 
\sum_{\ri \eta \in \cP} R_{ij}^{(\eta)} \frac{v \eta}{v \eta + \ri \omega} 
\e^{(t-t_0)(v \eta+\ri \omega)} \right ]\, , \quad t >t_0\, , 
\label{C2}
\end{equation} 
\end{widetext} 
which represents the off-critical generalization of the formula (\ref{C1}) we are looking for. 
Notice that away from criticality $G_{ij}(\omega, t-t_0)$ is in general complex, showing 
that off-critical junctions may have a nontrivial impedance. Since eq. (\ref{lrt2}) involves 
the commutator of the field $\ph$, the conductance does not depend on the parameter 
$\tau$, but depends on the time $t-t_0$ elapsed after switching on the external field. 

Before discussing the main features of (\ref{C2}), we would like to mention two 
useful checks. First of all we observe that at criticality, where scale invariance 
implies that $\S$ is constant ($k$-independent), 
the sum over the poles vanishes and (\ref{C2}) precisely reproduces the expression 
(\ref{C1}) from the introduction. A second highly nontrivial check is the Kirchhoff 
rule 
\begin{equation} 
\sum_{i=1}^n  G_{ij}(\omega, t-t_0) = 0 \, , 
\label{K5}
\end{equation}
which is a consequence of the conservation of the electric charge, namely 
of equations (\ref{K2}, \ref{K4}). 

The sum in the right hand side of (\ref{C2}) runs over both negative and 
positive poles $\cP = \cP_-\cup \cP_+$. 
The antibound states $\cP_-$ produce damped oscillations in $t-t_0$. 
If bound states are absent ($\cP_+ = \emptyset$),  
\begin{multline} 
\lim_{t\to \infty} G_{ij}(\omega, t-t_0) = 
\lim_{t_o\to -\infty} G_{ij}(\omega, t-t_0) = \\
G_{\rm line} \left [\delta_{ij} - S_{ij}\left (\frac{\omega}{v} \right )\right ] \, ,  
\label{limits}
\end{multline} 
which gives the conductance one will observe in the regime (i) long time after 
switching on the external field.  

{}Finally, the bound states $\cP_+$ give origin to oscillations 
whose amplitude is growing exponentially with $t-t_0$. 
The oscillations in (\ref{C2}) provide therefore a nice experimental 
signature for testing the analytic structure of the scattering matrix $\S(k)$.

\section{Outlook and conclusions} 

Off-critical Luttinger junctions are characterized by a scattering matrix $\S(k)$ which admits in general 
bound and antibound states. The presence/absence of bound states determines 
two different regimes of the theory. 
A scattering matrix without bound states gives raise to an isolated equilibrium 
system. One bound state is enough to change radically the situation. 
We have show in fact that each such a state generates an oscillator degree of freedom, 
whose contribution is fixed by local commutativity up to a common free parameter $\tau$. 
These additional degrees of freedom break the invariance under time translations 
and drive the system out of equilibrium. Accordingly, 
the vacuum energy is time dependent: it decays exponentially in the interval 
$(-\infty, \tau)$ and grows at the same rate in $(\tau, \infty )$. These two intervals are 
related by time reversal and are characterized by a nontrivial outgoing and incoming vacuum energy flows. 
Recalling that the traditional way \cite{CL} for driving a system out of equilibrium is to couple it 
with a ``bath" of external oscillators, we have shown above that under certain conditions such oscillators 
can be automatically generated by boundary effects. Junctions with bound states provide 
therefore an intrinsic mechanism for constructing non-equilibrium quantum systems,  
whose behavior is governed by purely boundary phenomena. 
Since this mechanism is based on the general physical requirement of 
causality, it extends\cite{BMS} also to systems in space dimensions greater then one. 

The off-critical conductance formula (\ref{C2}) has various interesting properties. 
A first remarkable feature of (\ref{C2}) is the different 
impact of antibound and bound states, which hopefully can be tested experimentally. 
The result (\ref{C2}) turns out to be useful also at the level of effective theory for quantum wire 
networks with several junctions, where it applies for the effective scattering matrix 
$\S_{\rm eff}(k)$ associated to the external edges of the network.

\acknowledgments
We thank Pasquale Calabrese, Benoit Dou\c{c}ot, In\`es Safi and Robert Schrader for 
enlightening discussions and correspondence. 
The research of B. B. has been supported in part by the NSF grant PHY-0757868.

\appendix 

\section{Effect of the bound states of $\S(k)$ on the field $\ph$} 
\label{appA} 

We assume below that $\S(k)$ admits bound states, the case 
$\cP_+ = \emptyset$ being standard\cite{Bellazzini:2006jb}. 
When $\cP_+ \not= \emptyset$  it is natural to represent 
the field $\ph$ as a linear combination 
\begin{equation} 
\ph (t,x,i) = \phs (t,x,i) + \phb (t,x,i) \, , 
\label{dec1}
\end{equation} 
where $\phs$ collects the contribution of the {\it scattering} states 
of $\S(k)$ and $\phb$, that of its {\it bound} states. 
The scattering component $\phs$ is known from previous 
studies \cite{Bellazzini:2006jb}. One has 
\begin{multline} 
\phs (t,x,i) = \\
\int_{-\infty}^{\infty} \frac{\rd k}{2\pi \sqrt
{2|k|}}
\left[a_i^\ast (k) \e^{\ri (|k|t-kx)} +
a_i (k) \e^{-\ri (|k|t-kx)}\right ] \,  , 
\label{sol1}
\end{multline} 
where $\{a_i(k),\, a^*_i(k)\, :\, k\in \RR\}$ generate the 
reflection-transmission algebra\cite{Mintchev:2002zd,Mintchev:2003ue,Mintchev:2004jy} $\A$ 
\begin{multline} 
[a_i(k)\, ,\, a_j(p)] = [a^*_i (k)\, ,\, a^*_j (p)] = 0 \, , \\
[a_i(k)\, ,\, a^*_j (p)] = 2\pi [\delta (k-p)\delta_{ij} + \S_{ij}(k)\delta(k+p)] \, , 
\label{rta}
\end{multline} 
and satisfy the constraints 
\begin{eqnarray} 
a_i(k) &=& \sum_{j=1}^n \S_{ij} (k) a_j (-k) \, , 
\label{constr1} \\ 
a^\ast_i (k) &=& \sum_{j=1}^n a^\ast_ j(-k) \S_{ji} (-k)\, ,    
\label{constr2}
\end{eqnarray} 
which are consistent because unitarity and Hermitian analyticity of $\S(k)$ imply 
$\S(k) \S(-k) = \II$. 

Since for $\cP_+ \not= \emptyset$ the scattering states are not complete,  
the scattering component $\phs$ is neither a canonical nor a local field. In fact, 
keeping in mind that $\xt_{12} >0$ and using the Cauchy integral formula one finds 
\begin{multline} 
[(\prt_t\phs )(0,x_1,i_1)\, ,\, \phs (0,x_2,i_2)] = \\
-\ri \delta_{i_1i_2}\, \delta (x_1-x_2) + 
\ri \sum_{\ri\eta \in \cP_+}\eta \e^{-\eta
\xt_{12}} R^{(\eta )}_{i_1i_2} \, , 
\label{etccr1}
\end{multline} 
and, at space-like separations $t_{12}^2-x_{12}^2 <0$, 
\begin{multline} 
[\phs (t_1,x_1,i_1)\, ,\, \phs (t_2,x_2,i_2)] = \\
\ri \sum_{\ri \eta \in \cP_+} \e^{-\eta \xt_{12}} \sinh (\eta t_{12})\, R^{(\eta)}_{i_1i_2} \, . 
\label{loc1}
\end{multline} 

We turn now to the construction of the boundary component $\phb$ 
which is designed in such a way that the total field (\ref{dec1}) is both canonical and local.
Let us introduce first the set $\I_+ = \{i\, :\, \eta_i > 0\}$ 
labeling the {\it different} poles of $\S(k)$ in the upper half plane. 
Then the normalizable solutions of the wave equation on 
$\Gamma$ corresponding to the bound states are $\{\e^{-\eta_i (x\pm t)}\, :\, i\in \I_+\}$. 
The main idea now is to associate with each index $i\in \I_+$ a quantum oscillator. 
The quantum boundary degrees of freedom are described therefore by the algebra $\cal B$ 
generated by $\{b_i, b^\ast_i\, :\, i\in \I_+\}$, which satisfy 
\begin{equation} 
[b_{i_1}\, ,\, b_{i_2}] =  
[b_{i_1}^\ast \, ,\, b_{i_2}^\ast ] = 0\, , \qquad  
[b_{i_1}\, ,\, b_{i_2}^\ast ] = 
\delta_{i_1 i_2} \, ,
\label{bccr2}
\end{equation} 
and commute with $\{a_i(k),\, a_i^\ast (k)\}$. Now, the field $\phb$ is defined by\cite{Bellazzini:2008cs}  
\begin{multline} 
\phb (t,x,i) = 
\frac{1}{\sqrt{2}} \sum_{j\in \I_+} \U_{ij}\bigl [\left (b^\ast_j + b_j\right ) \e^{-\eta_j (x+t -\tau)} 
+ \\ \ri \left (b^\ast_j - b_j \right ) \e^{-\eta_j (x-t+\tau)} \bigr ]\, , 
\label{bfield}
\end{multline} 
$\U$ being the unitary matrix which appears in (\ref{d1}). By construction 
the field (\ref{bfield}) satisfies the equation of motion (\ref{eqm1}), the boundary 
condition (\ref{bc}) and depends on the parameter $\tau \in \RR$, whose 
physical meaning has been discussed in section II.D. One can also verify\cite{Bellazzini:2008cs} that 
the total field (\ref{dec1}) is both canonical and local. In fact, 
\begin{multline} 
[\phb (t_1,x_1,i_1)\, ,\, \phb (t_2,x_2,i_2)] = \\
-\ri \sum_{\ri \eta \in \cP_+} \e^{-\eta \xt_{12}} \sinh (\eta t_{12})\, R^{(\eta)}_{i_1i_2} \, , 
\label{loc2}
\end{multline} 
which precisely compensates the right hand side of (\ref{loc1}). 

The Hamiltonian $H$, generating the time evolution of $\ph$ according to 
\begin{equation} 
[H\, ,\, \ph (t,x,i)] = -\ri (\partial_t\ph) (t,x,i) \, . 
\label{ham00}
\end{equation} 
is given by  
\begin{equation} 
H = H^{(s)} + H^{(b)} \, , 
\label{ham0} 
\end{equation}  
where the scattering contribution $H^{(s)}$ has the standard form 
\begin{equation}
H^{(s)} = \frac{1}{2}\sum_{i=1}^n\int_{-\infty}^{\infty} \frac{\rd k}{2\pi}
|k|\, a_i^\ast (k) a_i (k)\, , 
\label{ham1}
\end{equation}   
and 
\begin{equation}
H^{(b)} = \frac{\ri}{2}\sum_{j\in \I_+} \eta_j 
\left (b_j^2-b_j^{\ast 2}\right )\, .  
\label{ham3}
\end{equation}  
We see that $H^{(b)}$ is not at all the Hamiltonian of the conventional harmonic oscillator, 
which is the origin of the peculiar time dependence of the vacuum energy density (\ref{venergy}).  

For deriving the conductance it is convenient to plug in (\ref{lrt2}) the following integral representation of the 
commutator
\begin{widetext} 
\begin{equation}
[\ph(t_1,x_1,i_1)\, ,\, \ph(t_2,x_2,i_2)] = 
-\ri \int_{-\infty}^{\infty} \frac{\rd k}{2\pi} \frac{1}{k}\sin(kt_{12}) \left [\e^{\ri k x_{12}}\delta_{i_1i_2} + 
\e^{\ri k \tx_{12}}\S_{i_1i_2}(k) \right ]  - 
\ri \sum_{\ri \eta \in \cP_+} \e^{-\eta \xt_{12}} \sinh (\eta t_{12})\, R^{(\eta)}_{i_1i_2} \, . 
\label{comm}
\end{equation} 
\end{widetext} 
We stress that (\ref{comm}) is $\tau$-independent. Finally, the current-current correlator 
(\ref{jj}-\ref{dw}) is obtained in the Fock representation of the algebras 
$\A$ and $\cal B$, characterized as usual by a vacuum state $\Omega$ annihilated by 
$a_i(k)$ and $b_i$. 

In conclusion, we would like to stress once more the fundamental role played by 
causality (local commutativity) in the above construction. In fact, this deep physical 
requirement fixes (up to the parameter $\tau$) the whole structure of the field $\ph$ and 
its time evolution.

\vfill\eject 
%\newpage 


\begin{thebibliography}{99}
\bibitem{f1} 
Sometimes the combinations $g_{2,4}=2(g_{-}\mp g_{+})$ are used as well.
\bibitem{f4}
We stress that this is a purely boundary effect 
and that the energy momentum tensor in the bulk of $\Gamma$ 
satisfies the continuity equation.
\end{thebibliography}
\end{document}